\documentclass[fleqn,twoside]{article}
\usepackage{espcrc2}
\usepackage{epsfig,rotating}

\def\nue{\ensuremath{\nu_{e}}}
\def\nubare{\ensuremath{\overline{\nu}_{e}}}

\def\numu{\ensuremath{\nu_{\mu}\ }}
\def\nubarmu{\ensuremath{\overline{\nu}_{\mu}}}

\newcommand{\dmtt}{\ensuremath{\Delta m^2_{23} \,}}

\newcommand{\He}{\ensuremath{^6{\mathrm{He}\,}}}
\newcommand{\Ne}{\ensuremath{^{18}{\mathrm{Ne}\,}}}

\newcommand{\thetaot}{\ensuremath{\theta_{13}}\,}
\newcommand{\thetatt}{\ensuremath{\theta_{23}}\,}

\newcommand{\sigdm}{\ensuremath{{\rm sign}(\Delta m^2_{23})\ }}
\newcommand{\delCP}{\ensuremath{\delta_{\rm CP}\ }}
\begin{document}
 \textfloatsep 12pt
 \setcounter{dbltopnumber}{2}    
 \renewcommand{\dbltopfraction}{0.9}	
 \renewcommand{\textfraction}{0.07}	
 \renewcommand{\floatpagefraction}{0.8}	
 \renewcommand{\dblfloatpagefraction}{0.8}	
\title{\vspace*{-0.5cm}Physics Potential of the $\gamma=100,100$ Beta Beam }
\author{Mauro Mezzetto
\address{Istituto Nazionale Fisica Nucleare, Sezione di Padova.
Via Marzolo 8, 35100 Padova, Italy. 
}}
\begin{abstract} 
The physics potential of a Beta Beam fired from CERN to a 440 kton
water Cerenkov detector at a distance of 130 Km is computed.
\end{abstract}
\maketitle
%
%
%
\section{Introduction}
Beta Beam ($\beta$B) \cite{Piero} performances 
 have been computed for $\gamma(\He)=$ 66 \cite{beta}, 100 \cite{MyNufact04} \cite{Donini:2004hu} \cite{JJHigh2},
150 \cite{JJHigh2},200 \cite{LindnerBB},350 \cite{JJHigh2}, 500 \cite{JJHigh1} \cite{LindnerBB},
1000 \cite{LindnerBB}, 2000 \cite{JJHigh1}, 2488 \cite{Terranova}.
A review can be found in \cite{MyNeutrino04},
physics potential of low gamma $\beta$B has been studied in \cite{Volpe}.
Performances of $\beta$B with $\gamma>150$ are extremely promising but
 rather speculative, because they are not based
on an existing accelerator complex nor on a robust estimation of the ion decay rates.
For a CERN based Beta Beam, fluxes have been estimated
in \cite{Lindroos}, and the physics potential for a beam fired to a
440 kton water Cerenkov detector \cite{UNO} hosted under the Frejus, at 130 km from CERN,
has been firstly computed in \cite{beta}. So far the Frejus site \cite{Mosca} 
is the only realistic
candidate at a suitable baseline for a CERN based $\beta$B with
 $\gamma \leq 150$.

This work updates and consolidates performances computed in
 \cite{MyNufact04} for a $\beta$B where both \He\ and \Ne\  ions have
$\gamma=100$ ($\beta$B100), the optimal setup for a 130 km distance.
In particular signal efficiency, background
estimation, energy binning will be revised and all the sensitivities will be computed using the
open source program Globes \cite{Globes}, allowing for an
 estimation of performances
in presence of degeneracies.
 Finally the Beta Beam performances will be
computed as function of the duty cycle.
%
\section{Signal and backgrounds}
\subsection{Signal efficiency}
\label{sec:eff}
\begin{figure}
    \vspace*{-0.6cm}
    \hspace*{-0.5cm}
    \epsfig{file=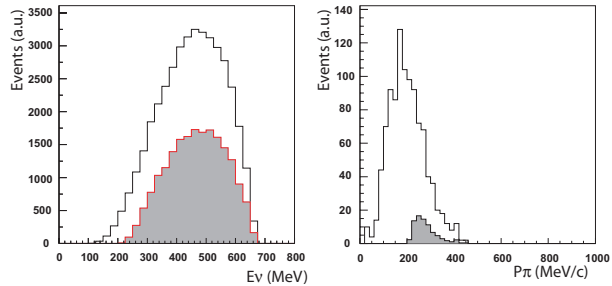,width=0.5\textwidth} 
    \vspace*{-1.5cm}
    \caption{Left: Event reduction for \Ne\ oscillated events (left) and 
    pion background, $\pi^+ + \pi^-$ (right).}
    \label{fig:ratio_eff}
\end{figure}

Signal events in $\beta$B, \numu charged current (CC) events, are selected with standard 
SuperKamiokande particle identification algorithms.
The muon identification is reinforced by asking for
the detection of the Michel decay electron. 
Data reduction is shown in fig.~\ref{fig:ratio_eff} for \Ne\  events and
detailed in tab.~1.
%
\subsection{Energy binning}
\begin{figure}[htb]
    \vspace*{-0.5cm}
  \epsfig{file=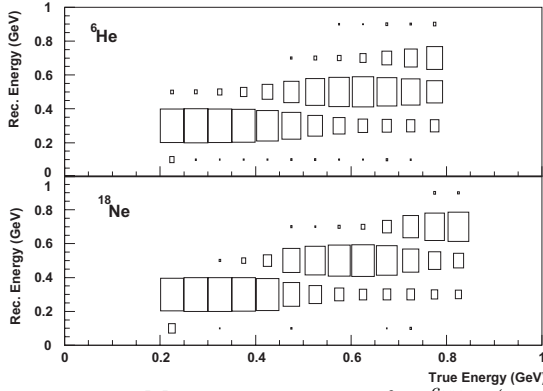,width=0.45\textwidth} 
    \vspace*{-1.1cm}
    \caption{Migration matrices for \He\ (upper plot) and \Ne\ (lower plot).}
    \label{fig:migmat}
\end{figure}

As pointed out in reference \cite{JJHigh2},
 it is necessary to use a
migration matrix to properly handle the Fermi motion smearing in
the $\beta$B100 energy range.
The matrices, computed with Nuance \cite{Nuance}, 
have 25 true energy bins and 5 reconstructed energy bins in the energy range
$0 < E_\nu < 1$ GeV, see fig.~\ref{fig:migmat}.
 The migration matrix approximation
has a visible effect in the Leptonic CP Violation  
discovery potential,
 as discussed in section \ref{sec:sensitivities}.
%
\subsection{Atmospheric neutrino backgrounds}
Atmospheric neutrino can constitute an important source of
backgrounds \cite{Piero}. They can be suppressed only by keeping
a very short duty cycle, and this in turn is one of the most
challenging bounds on the design of the Beta Beam complex.

To compute the rate of this background, the total integrated rate of
atmospheric
\nue\  in SK, in the energy range of $\beta$B100, is considered.
This rate is scaled to the 440 kton fiducial mass and corrected
 for the difference
of atmospheric neutrino fluxes between Kamioka and Frejus \cite{Barr}.
The direction of the incident neutrino
at 400 MeV, according to  Nuance, can be reconstructed with a
$\sim 0.25$ radians resolution. A $2\sigma$ cut around the CERN-Frejus
direction is then applied. Finally  the efficiency curves
of section~\ref{sec:eff} are applied to the remaining events. 
Assuming 
8 bunches 6.25 ns long in the 7 km long
decay ring, the duty cycle is $2.2 \cdot 10^{-3} $.
Atmospheric neutrino backgrounds sum up to
13 background events per ion specie in a 4400 kton-year exposure ($10^7$ s/year).

In the following, sensitivities are computed with this background,
 the effect of a higher duty cycle is discussed in
session \ref{sec:dutycicle}.
\subsection{Charged pions background}
\label{sec:pions}
Charged pions generated in NC events (or in NC-like events where
the leading muon goes undetected) are the main source of background
for the experiment.
To compute this background 
inclusive NC and CC events have been generated with the $\beta$B100 spectrum.
Events have been selected where the only visible track is a charged pion
above the Cerenkov threshold. Particle identification efficiencies
 have been applied to those particles.
With Geant 3.21 the probability for a pion to survive
in water until its decay have been computed
\footnote{cross-checked 
with a Fluka 2003 simulation \cite{Campagne}}.
This probability is different for positive and
negative pions, these latter having a higher probability to be
absorbed before decaying.
The surviving events are backgrounds and the reconstructed
neutrino energy is computed mis-identifying these pions as
muons.
Event rates are reported in tab.~1.
\begin{table}
     \vspace*{-0.5cm}
     \label{tab:sigbck}
     \caption{Events in a 4400 kton-year exposure.
     \numu(\nubarmu) CC events 
     are computed assuming full oscillation, pion backgrounds are computed
     from \nue(\nubare) CC+NC events.} 
     \renewcommand{\arraystretch}{0.9}
     \hspace*{-0.2cm}
     \begin{tabular}{|c|rrr|rrr|}
      \hline
       & \multicolumn{3}{ c }{Ne18} & \multicolumn{3}{|c|}{He6} \\
      \hline
             & \numu CC & $\pi^+$ & $\pi^-$ & \nubarmu CC & $\pi^+$ & $\pi^-$ \\
      \hline
      in & 139181     &  863   &  561 & 107571     &  952   &  819 \\
      pid & 105923    &  209   &  123 & 83419    &  242   &  170 \\
      dcy      & 67888    &  103    &    6 & 67727    &  117   &    7 \\
      \hline
    \end{tabular}
\end{table}
These background rates are significantly smaller from what quoted in
\cite{MyNufact04}, where pion decays  were computed with the
same probabilities of the muons.
%
\section{Sensitivities}
\label{sec:sensitivities}
  Sensitivity to \thetaot is by definition the performance of the experiment
  in absence of signal. 
  It has been 
  computed for 10 years running time (5 \He + 5 \Ne) with $5.8 \cdot 10^{18}$
  useful \He\  decays/year
  and $2.2 \cdot 10^{18}$ useful \Ne\  decays/year.
  Appearance and disappearance channels have been combined together.
  Input values are $\thetatt=\pi/4$, $\dmtt=2.5 \cdot 10^{-3}$ eV$^2$,
  $\sin^2{\theta_{12}}=0.315$, $\Delta m^2_{12}=7.9 \cdot 10^{-5}$ eV$^2$,
 \sigdm=+1.
  Parameter errors have been fixed to the T2K sensitivities for the
atmospheric parameters \cite{T2K} and to the present values of the solar parameters:
 $\thetatt=5\%  $,  $\dmtt=4\% $,
 ${\theta_{12}}=10\%$, $\Delta m^2_{12}=4\%$.
  The systematic errors for
      signal and backgrounds has been fixed to 2\%.

   The sensitivity curve, fig.~3, is a 6 parameters fit minimized over the solar and the
   atmospheric parameters and projected
  over \thetaot.
  Degeneracies induced by the unknown values of \sigdm and \thetatt
  are not  accounted for.
 \begin{figure}[b]
    \vspace*{-0.2cm}
   {\epsfig{file=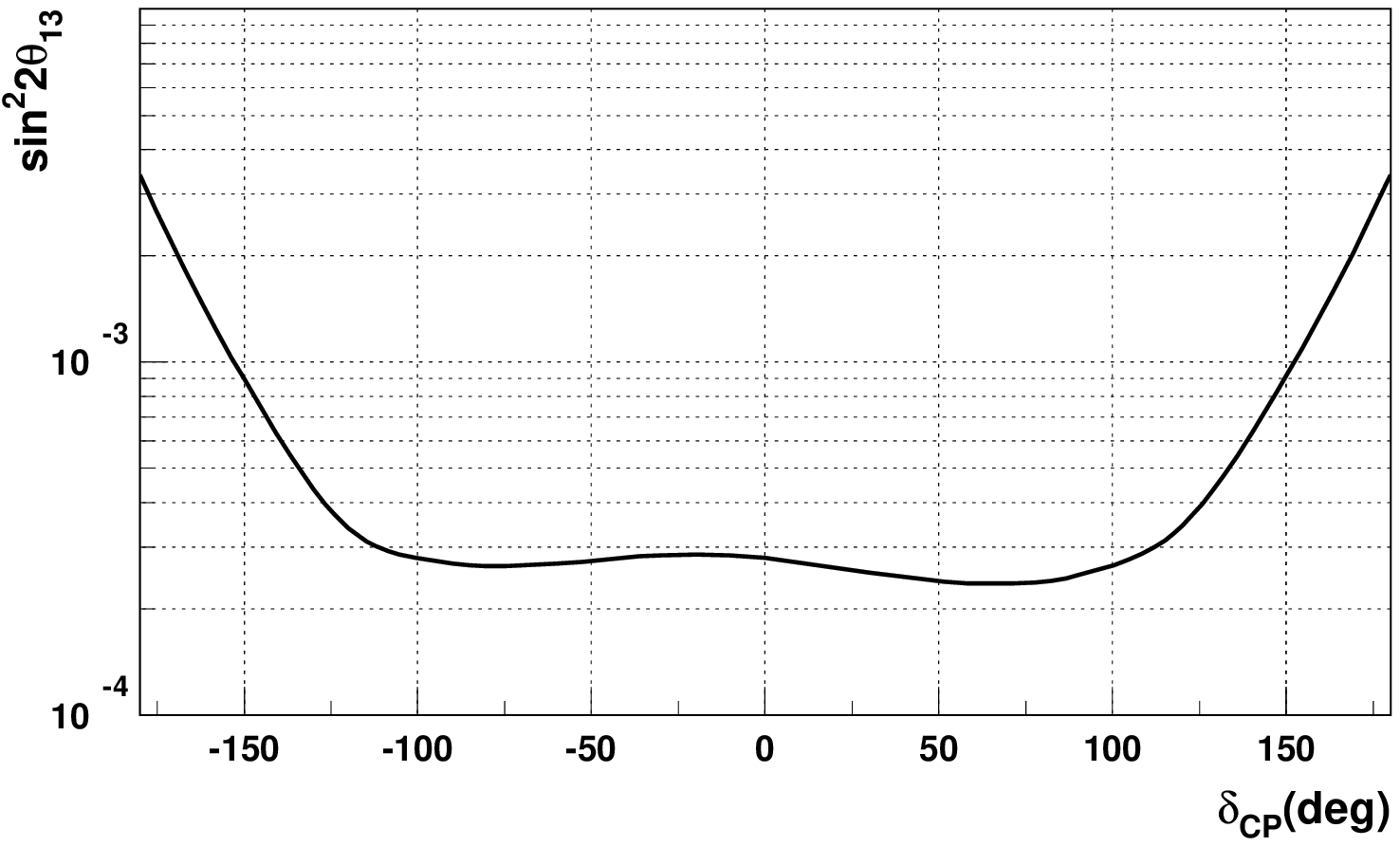,width=.45\textwidth}}
    \vspace*{-1.1cm}
   \label{fig:th13}
   \caption{$\thetaot$ sensitivity at 90\% CL ($\Delta \chi^2 > 4.61$) as function
   of \delCP (see text).}
 \end{figure}

 In case of signal it is important
to quantitatively assess the discovery potential for
 leptonic CP violation (LCPV).
 It is computed at $3 \sigma$ ($\Delta \chi^2= 9.0$)
 taking into account all the parameter errors and all
 the possible degeneracies. As common practice in
 literature $\theta_{23}=40^\circ$
 has been used, to leave room for the octant ($\pi/4-\thetatt$) degeneracy.
 In fig.~4 discovery potential is computed under 4 different hypotheses
 of the true parameters,
 normal: \sigdm=1, $\thetatt<\pi/4$;
 sign: \sigdm=-1, $\thetatt<\pi/4$; 
 octant: \sigdm=1, $\thetatt>\pi/4$;
 mixed: \sigdm=-1, $\thetatt>\pi/4$.
 Each of these 4 true values  combinations 
 has been fitted with the 4 possible fit combinations of \sigdm and \thetatt.
 Also shown are the LCPV discovery potentials
 neglecting the degenerate solutions.
 Effect of degeneracies are sometimes visible for high values
 \begin{figure}[h]
    \vspace*{-0.7cm}
   \epsfig{file=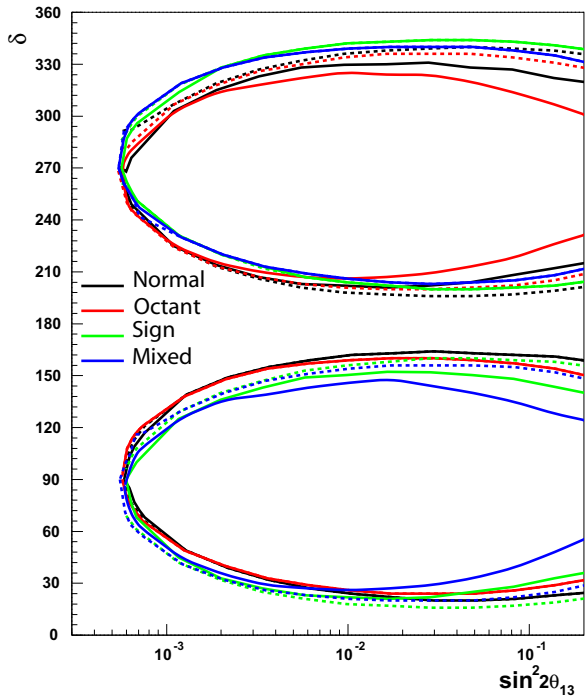,height=.46\textheight}
    \vspace*{-1.3cm}
   \label{fig:deltacp}
   \caption{LCPV discovery potential at $3\sigma (\Delta\chi^2>9.0)$
   computed for the 4 different options about the true values of
   \sigdm and \thetatt\  (see text). 
   Dotted curves are computed neglecting the effects of the clone
   solutions.}
 \end{figure}
 of \thetaot, precisely the region
 where they can be reduced by a combined analysis with atmospheric neutrinos
 \cite{Schwetz}.

Several effects play significant roles in the final LCPV discovery potential,
as
gaussian approximation for the energy binning or
systematic errors bigger than 2\%,
  as shown in fig.~5.
 \begin{figure}[t]
    \vspace*{-0.6cm}
   \epsfig{file=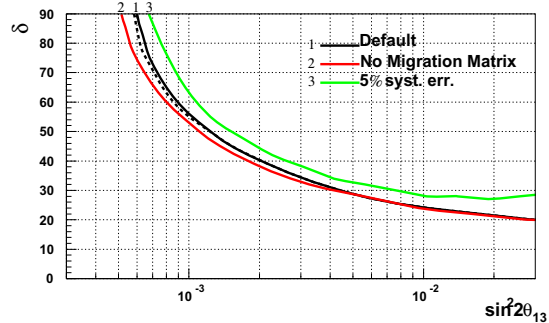,width=.45\textwidth}
    \vspace*{-1.1cm}
   \label{fig:diff_effects}
   \caption{LCPV discovery potential at $3\sigma 
   (\Delta\chi^2>9.0)$,
   under three different options of the input parameters
   (see text) and computed without
   the degenerate solutions (dotted line).  }
 \end{figure}
\section{Duty cycle}
\label{sec:dutycicle}
A critical parameter in designing the Beta Beam complex is the duty cycle.
Due to the available longitudinal acceptance in the decay ring,
$\beta B$ fluxes can only be increased by increasing the number of bunches
in the decay ring \cite{MatsFluxes}.
An extremely short duty cycle, aimed to keep the atmospheric background close
to zero, was the choice for the $\gamma=66$ $\beta$B \cite{beta}, 
where the experimental 
backgrounds were close to zero. For the $\gamma=100$ option, a modest rate
of atmospheric neutrino backgrounds is tolerable. To assess the highest possible
duty cycle for $\beta B 100$,
 LCPV performances have been computed (neglecting degeneracies) by varying the duty cycle under two 
hypotheses: the overall fluxes remain constant or the overall fluxes rise as
function of the
duty cycle as discussed in \cite{MatsFluxes} (fig.~6).
 \begin{figure}
    \vspace*{-0.3cm}
      \epsfig{file=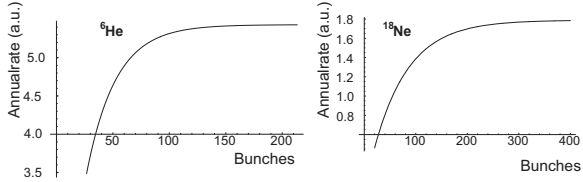,width=.48\textwidth}
    \vspace*{-1.4cm}
   \label{fig:dc}
   \caption{Ion decay rates (a.u) as function of the duty cycle for \He\  (left)
    and \Ne\  (right). Taken from reference \cite{MatsFluxes}.}
 \end{figure}

Fig.~7 right, shows performances at constant fluxes.
The degradation of performances is evident at the smaller values of 
\thetaot, where the background level is the dominant factor, while at
higher values of \thetaot the performances are rather stable. 
The left plot of fig.~7 shows 
performances computed with the flux-duty cycle relationship of fig.~6.
Here the nominal fluxes have been assumed for 20 bunches in the decay ring
both for \He\  and \Ne\  ions. 
This assumption is more optimistic of what quoted in \cite{MatsFluxes},
the purpose of this plot is not to display performances of $\beta$B100 but to
show how performances  scale as function of the duty cycle.
Discovery potential increases up the point where a flux saturation
occurs, roughly for 75 \He\  bunches  and for 150 \Ne\  bunches.
This latter plot shows that
 the high intensity frontier for the Beta Beams
is as promising as the high gamma scenarios.
 \begin{figure}
    \vspace*{-1.4cm}
    \epsfig{file=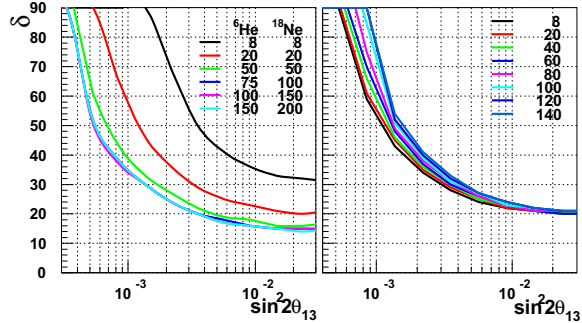,width=.48\textwidth}
    \vspace*{-1.4cm}
   \label{fig:ppvsdc}
   \caption{LCPV discovery potential for different values of 
   the number of ion bunches
   circulating in the decay ring computed following
   the curves of fig.~6 (left) or keeping
   constant the ion decay rates (right).}
 \end{figure}
%
\section{Conclusions}
The Beta Beam design study for the baseline CERN option is running.      
It will provide the final estimation of neutrino rates and possible
pathways to increase the performances. The physics potential, as can be
predicted today, shows that it  is worth the effort. \\

I'm very grateful to J.E.~Campagne, P.~Huber, M.~Lindroos, M.~Maltoni and
T.~Schwetz for the many illuminating discussions during the preparation of
these studies.


\begin{thebibliography}{99}
%
\bibitem{Piero}
 P. Zucchelli, Phys.\ Lett.\ B {\bf 532} (2002) 166.
%
\bibitem{beta}
 M. Mezzetto,
  J.Phys.G29:1781-1784, 2003; [hep-ex/0302007].
 J.~Bouchez, M.~Lindroos and M.~Mezzetto,
 AIP conference proceedings, Vol. 721, 37-47, 2003.
 [hep-ex/0310059].
%
\bibitem{MyNufact04}
  M.~Mezzetto,
  Nucl.\ Phys.\ Proc.\ Suppl.\  {\bf 149} (2005) 179.
%
\bibitem{Donini:2004hu}
 A.~Donini { et al.},
  Nucl.\ Phys.\ B {\bf 710} (2005) 402
 [arXiv:hep-ph/0406132].
%
\bibitem{JJHigh2} 
 J.~Burguet-Castell { et al.},
  arXiv:hep-ph/0503021.
%
\bibitem{JJHigh1}
 J.~Burguet-Castell { et al.},
 Nucl.\ Phys.\ B {\bf 695} (2004) 217
 [arXiv:hep-ph/0312068].
%
%
\bibitem{LindnerBB}
  P.~Huber { et al.},
  arXiv:hep-ph/0506237.
%
\bibitem{Terranova}
 F.~Terranova {et al.},
 Eur.\ Phys.\ J.\ C {\bf 38} (2004) 69
 [arXiv:hep-ph/0405081].
%
\bibitem{MyNeutrino04}
  M.~Mezzetto, 
  Nucl.\ Phys.\ Proc.\ Suppl.\  {\bf 143} (2005) 309
  [arXiv:hep-ex/0410083].
%
\bibitem{Volpe}
 C.~Volpe,
  J.\ Phys.\ G {\bf 30} (2004) L1
  [arXiv:hep-ph/0303222].
\bibitem{Lindroos}
 B.~Autin { et al.},
 physics/0306106. 
 M. Benedikt, S. Hancock and M. Lindroos,
 Proceedings of EPAC, 2004,
 http://accelconf.web.cern.ch/AccelConf/e04.
 M. Lindroos, proceedings of this conference.
\bibitem{UNO}
  UNO Collaboration, hep-ex/0005046.
%
\bibitem{Mosca}
  L.~Mosca,
  Nucl.\ Phys.\ Proc.\ Suppl.\  {\bf 138} (2005) 203.
%
\bibitem{Globes}
  P.~Huber, M.~Lindner and W.~Winter,
  Comput.\ Phys.\ Commun.\  {\bf 167} (2005) 195
  [arXiv:hep-ph/0407333].
%
\bibitem{Nuance}
D.~Casper,
  Nucl.\ Phys.\ Proc.\ Suppl.\  {\bf 112} (2002) 161
  [arXiv:hep-ph/0208030].
%
\bibitem{Barr}
G.~Barr, T.~K.~Gaisser and T.~Stanev,
  Phys.\ Rev.\ D {\bf 39} (1989) 3532.
%
\bibitem{Campagne}
J.E.~Campagne, private comunication.
%
\bibitem{T2K}
Y.~Itow { et al.},
 hep-ex/0106019.
\bibitem{Schwetz}
 P.~Huber, M.~Maltoni and T.~Schwetz,
  Phys.\ Rev.\ D {\bf 71} (2005) 053006
  [arXiv:hep-ph/0501037].
%
\bibitem{MatsFluxes}
M. Lindroos, EURISOL DS/TASK12/TN-05-02.
 M. Benedikt, A. Fabich, S. Hancock and M. Lindroos,
proceedings of this conference.
%
\end{thebibliography}
\end{document}